# Sign changing pairing in single layer $FeSe/SrTiO_3$ revealed by nonmagnetic impurity bound states

Huimin Zhang[1], Zhuozhi Ge[1,2], Michael Weinert[2], and Lian Li[1★]

[1]Department of Physics and Astronomy, West Virginia University, Morgantown, WV 26506, USA

[2]Department of Physics, University of Wisconsin, Milwaukee, WI 53201, USA

[★]Correspondence to: lian.li@mail.wvu.edu

## Abstract

The discovery of high-temperature superconductivity in single layer FeSe epitaxially grown on $SrTiO_3$(001) substrates has instigated extensive debate over whether its pairing symmetry is conventional sign-preserving *s*-wave or unconventional sign-changing. Here we probe the pairing state for single layer $FeSe/SrTiO_3$ grown by molecular beam epitaxy using scanning tunneling microscopy/spectroscopy. We observe robust in-gap bound states induced by non-magnetic Fe-vacancy defects within the Fe-plane, which exhibit strong spatial electron-hole asymmetry with two-fold symmetry in hole states and four-fold in electron states. The bound states exhibit no energy shift or splitting under an applied magnetic field, consistent with a sign-changing order parameter. This is further confirmed by defect bound state quasiparticle interference that shows a sign changing behavior with a pair of corresponding peaks at the positive and negative energies near the impurity bound states. Our findings provide unambiguous evidence for a sign-changing pairing symmetry for single layer $FeSe/SrTiO_3$.

## Introduction

The Cooper pairs in a superconductor can be locally broken by impurity scattering, leading to Bogoliubov quasiparticle excitations that can reveal the nature of the superconducting pairing state[1]. For conventional superconductors, only magnetic impurities can break the Cooper pairs, while for unconventional superconductors, both magnetic and non-magnetic impurities can induce such bound states. Hence, probing bound states and quasi-particle scattering at the atomic scale using scanning tunneling microscopy/spectroscopy (STM/S) has been a proven phase sensitive method for probing pairing symmetry of superconductors. Well known examples include the observation of asymmetric spatial distribution of bound states for electrons and holes in cuprates, which provides unambiguous evidence for $d_{x^2-y^2}$ wave pairing[1–4].

For Fe based superconductors, the determination of pairing symmetry has been challenging due to the multiple bands crossing the Fermi level, giving rise to a number of pairing symmetries[5]. The recent discovery of high-temperature superconductivity in single layer FeSe epitaxially grown on $SrTiO_3$(001) substrate[6,7] provides a model system to probe the pairing states of Fe-based superconductors. Because of electron doping from the STO substrate[8], the Fermi surface of FeSe/STO consists of only electron pockets at the Brillouin zone (BZ) corner *M* points, while the hole pockets at $\Gamma$ point are entirely below the Fermi energy[9,10]. The lack of hole pockets challenges the spin fluctuation picture that involves a sign reversal order parameter between the electron and hole pockets[11,12]. This has instigated extensive debate on the symmetry of pairing state in single layer FeSe, with leading contender being sign-preserving conventional *s*-wave pairing[13], and unconventional sign-changing extended *s*-wave[14,15] and nodeless *d*-wave pairing[16–19], where experimental validation are sorely needed[20].

Impurity scattering has been utilized to investigate the superconducting pairing symmetry of single layer FeSe/STO, however with conflicting results[13,21,22]. A plain *s*-wave has been suggested in an earlier STM work[13], based on quasiparticle interference off *adsorbates on the surface* of FeSe, where no impurity-induced bound states were observed for nonmagnetic Zn, Ag, and K adatoms. However, more recent work shows that even for nonmagnetic surface K[21] and Pb adatoms[22], bound states near the superconducting gap edge were found after careful data analysis. The emergence of the induced bound states by these nonmagnetic surface adatoms[21,22] indicates sign-reversing pairing scenarios, challenging the plain *s*-wave pairing symmetry. One concern for scattering by surface adatoms is that the induced bound states are located near the superconducting gap edges due to weak scattering[23], which maybe challenging to discern due to their close proximity to the coherence peaks. As such, the pairing symmetry for single layer FeSe/STO remains controversial, and unambiguous verification of induced in-gap bound states and quasiparticle scattering off *nonmagnetic impurities within the Fe plane* is crucial.

In this work, we probe bound states induced by non-magnetic impurities within the Fe-plane of single layer FeSe/STO, whose nonmagnetic nature is verified by magnetic field dependent studies. We observe robust in-gap bound states with strong spatial electron-hole asymmetry, which provides unambiguous evidence for sign-changing pairing symmetry. We further carry out systematic impurity scattering studies and apply the recently proposed defect bound state (DBS) quasiparticle interference (QPI) method[24–27] to provide further support for a sign reversal pairing order parameter, consistent with extended *s*-wave[14,15] or nodeless *d*-wave pairing[16].

## Results

**Non-magnetic defects in single layer FeSe/STO.** Figure 1**a** shows typical surface morphology of superconducting single layer FeSe film that is conformal to the step-terrace morphology of the STO substrate. The film is also characterized by a high density of grain boundaries (GBs), appearing as a network of meandering lines with low contrast, similar to previous reports[6,7]. Tunneling spectroscopy reveals a typical “U” shape superconducting gap structure with two coherence peaks, which exhibit spatial variations (Supplementary Fig. S1). Fitting the histogram of the superconducting gaps by Gaussian functions yields an average value of $_1$=11.1±0.1 meV and $_2$=17.5±0.1 meV, comparable to previous reports[6,13,21]. The superconducting transition temperature ($T_c$) is ~60 K from our own angle resolved photoemission spectroscopy (ARPES) measurements[28], comparable to 65 K in previous reports[10].

While defect density on as-grown single layer FeSe/STO film is low, at ~$5.0\times10^{11}$ $cm^{-2}$, higher density (~$1.5\times10^{13}$ $cm^{-2}$) can be introduced by capping the FeSe films with amorphous Se film first, and then annealing to remove the Se layer (Supplementary Fig. S2). Five types of defects can be created that are either two- or four-fold symmetric (Supplementary Fig. S3). Accounting for 90% of all defects are of types A and B (Figs. 1**c**&**d**), which are centered at the bridge site, i.e., Fe site, of the top-layer Se lattice, and appear as a dumbbell consisting of two bright lobes that are asymmetric in intensity at adjacent Se sites[29,30]. These dumbbell features have been reported in earlier studies and attributed to Fe vacancy and $Se_{Fe}$ substitutional defects. While the nature of the dumbbell defects is still under debate[29,21], comparison between our density functional theory calculations and STM images taken at liquid nitrogen temperature indicates that type A defects are Fe vacancies (Supplementary Fig. S4). In contrast, the four-fold symmetric defects (Supplementary Figs. S3**c-e**) are located at Se sites, likely Se-site related defects[21], and therefore are not the focus of this work.

Tunneling spectra taken at the center of dumbbell defects show strongly suppressed coherence peaks as well as pronounced resonance peaks inside the superconducting gap, as depicted in Figs.

1**e**&**f**. The energy position of bound states for Type A and B defects are −2.7 meV (inside the superconducting gap) and −8.3 meV (near the inner coherence peak), respectively. The resonance peak at the center of the impurity site can be up to three times stronger than the normal state conductance. Magnetic field dependent measurements (Figs. 1**e**&**f**) further indicate that while the intensities of these states are slightly suppressed in an external magnetic field; no shift or splitting of these peaks are observed up to 9 T out-of-plane or 2 T in-plane fields, indicating both these types of defects are nonmagnetic.

We have also carried out tunneling spectroscopy measurements at 400 mK with improved energy resolution of 0.36 meV, as determined by measuring the superconducting gap of a well-known superconductor Pb (Supplementary Fig. S5), which should be sufficient to resolve the 1 meV Zeeman splitting that's typical for this material system[31]. Magnetic-field dependent tunneling spectra of the dumbbell defects taken at this temperature are shown in Supplementary Fig. S6, where again no shifting or splitting of the bound states are observed at ± 9 T magnetic field. Note that while some fluctuations in the energy position of the bound state are observed with magnetic field, they are within the spectral energy resolution, and do not show a linear dependence on the magnetic field as would be expected from Zeeman effect. These results show unambiguously that the dumbbell defects, both type A and type B, are not magnetic.

We further note that these observations are in marked contrast to magnetic Fe vacancy defects in $KFe_2Se_2$ films[31], where the Zeeman shift of ~ 1 meV (at 9 T) is resolved at 400 mK, as well as a clear linear magnetic field dependence of the energy position of the bound states.

**Spatial and energy dependence of bound states.** Differential conductance maps of Type A dumbbell defect (Fig. 2**a**) further reveal that these in-gap bound states are confined to the vicinity of the defects in real space, but do exhibit a different spatial distribution for the electron- and hole-like states (Figs. 2**b**&**c**). Specifically, the hole-like (negative energy) bound state at –4.4 meV exhibits a two-fold symmetry (Fig. 2**b**), where the intensity is the strongest directly at the impurity site (Fig. 2**e**). In contrast, the electron-like (positive energy) bound state at +4.4 meV exhibits a four-fold symmetry with the highest intensity at the four lobes (Fig. 2**c**), as shown in the *dI*/*dV* spectrum taken at one of the lobes in Fig. 2**f**. A strong electron-hole asymmetry is clearly seen here where the local maxima and minima occur at different lattice sites with respect to the impurity. A detailed site dependent measurement of the bound states is shown in Supplementary Fig. S7, where the intensity of the hole state decreases from the defect site to the nearest neighbor (NN), next nearest neighbor (NNN), $3^{rd}$ NN, and so on, while the electron state emerges, with increasing intensity in the reverse order, reaching maximum at $3^{rd}$ NN. Overall, the spatial distribution at +4.4 meV shows a $C_2$ symmetry, while at –4.4 meV it is approximately $C_4$ symmetric. Moreover, the $C_2$ symmetry is along the dumbbell defect direction (Fig. 2**b**), whereas the relatively bright four-fold region is along the Fe-Fe direction (Fig. 2**c**), 45° to the Se-Se lattice, consistent with that they are within the superconducting Fe plane. Similar electron-hole asymmetry is also observed for type B defect, as shown in Supplementary Fig. S8. These observations of the in-gap bound states and spatial electron-hole asymmetry are similar to those of in *d*-wave cuprate superconductors[1–3], indicating an unconventional pairing symmetry in single layer FeSe/STO. Note that measurements at 77 K indicate that there are no impurity states within +/- 45 meV of the Fermi level for type A defects (Supplementary Fig. S9**b**). Furthermore, dI/dV mapping between ± 20 meV exhibit no electron-hole asymmetry (Supplementary S9**c**-**n**) in the normal state.

The presence of these defects, however, modifies the normal states around –0.26 eV. For type A, the peak shifts towards the Fermi level by as much as 20 meV (Supplementary Fig. S10**b**), indicating local hole doping. In contrast, for type B defects, the peak moves away from the Fermi level (Supplementary Fig. S10**d**), consistent with electron doping[8,32]. Interestingly, these shifts in the normal states are also directly correlated with the energy positions of the bound states. For example, as shown in Figs. 3**d**&**e**, for the three type A defects I, II, and III (marked in Fig. 1**b** and Figs. 3**a**-**c**), the bound states shift towards the Fermi level in tandem with the normal state.. Analysis of up to 70 defects indicates that the bound state energy position can vary between −1.7 and −7.5 meV. Plotted in Figs. 3**f**&**g** are histogram of the bound states, and the correlation between the normal-state and the bound-state peak energies. While the origin of such correlation is not known at the moment, the normal-state peak energy shifts downward as the bound state shifts towards the superconducting edge, suggesting that local electron doping is one of possible causes for the energy shift of the bound states. Note that this behavior deviates from that of LiFeAs, where the bound states induced by different impurities are pinned near the coherence peaks[33].

These observations of in-gap bound states *induced by non-magnetic impurities within the Fe plane*, particularly their spatial electron-hole asymmetry, i.e., a two-fold electron state and four-fold hole state, are the defining signatures of Bogoliubov quasiparticle excitations. In turn they strongly suggest a sign-changing unconventional order parameter in single layer FeSe/STO, which is further confirmed by the defect bound state phase-reference QPI discussed below.

**Defect bound state phase-reference quasiparticle interference.** Impurities, particularly those located within the Fe plane such as Fe vacancies can serve as strong scattering centers for quasiparticles and produce standing waves in real space, which can be better seen in the energy dependent conductance maps in Figs. 4**b**-**d** (topography image shown in Fig. 4**a**). The corresponding fast Fourier transforms (FFTs) are shown in Figs. 4**f**-**h**. Additional data at different energies are presented in Supplementary Figs. S11 and S12. The FFTs are fourfold symmetrized and subtracted by a 2D-gaussian-function background around $\Gamma$ point to improve signal-to-noise ratio. Typical nine-ring features in $q$ space are observed, similar to previous reports[13,34]. Based on Fermi surface maps obtained from ARPES[9,10], also sketched here in Fig. 4**e**, the quasiparticle scattering includes intra-pocket scattering $\mathbf{q_1}$ and inter-pocket scattering $\mathbf{q_2}$ and $\mathbf{q_3}$ as labeled in Fig. 4**f**. The scattering rings induced by $\mathbf{q_2}$ and $\mathbf{q_3}$ appear more as arcs, lacking four-fold symmetry, likely related to orbital features[13].

While these FFTs yield the intensities of scattering channels, they do not provide any phase information. For unconventional superconductors with sign change between different Fermi surfaces, additional phase shifts are expected near the bound state induced by nonmagnetic impurities. This phase information can be obtained from the recently developed defect bound states-QPI technique[24–27], where a pair of positive and negative peaks are expected to appear in $g_{pr}(q,+E)$ and $g_{pr}(q,-E)$, respectively, at energies near the nonmagnetic impurity bound-state energies $\pm E_B$ (see Methods).

As shown in Figure 5, a series of phase-corrected QPI maps are calculated for a type A defect (located at the center) at ±4.4 meV (bound state energy) and ±10.4 meV (coherence peak of $\Delta_1$), following equations (1)-(3). Additional data at different energies are presented in Supplementary Figs. S13 and S14. For positive energies (Figs. 5**e**&**f**), $g_{pr}(q,E)$ are positive everywhere, as expected. On the other hand, at negative energies (Figs. 5**g**&**h**), both positive and negative values are observed. Particularly for $E = -4.4$ meV, the region between the two circles (defined by $0.24\pi/a < \mathbf{q} < 0.64\pi/a$, where $a$ is the Fe-Fe bond length) is overwhelmingly negative, indicating that this pocket contains a sign-reversal inter- or intra-pocket scattering. The extent of this region changes with energy, which can be better illustrated by integrating the signal within the area as a function of energy, as shown in Fig. 5**i**. Clearly, a positive and negative peak is seen at + and − 4.4 meV, energy positions of the impurity-induced bound states, indicating a sign changing pairing parameter for single layer FeSe/STO. We note that this defect bound states -QPI method was also applied to $(Li_{1-x}Fe_x)OHFe_{1-y}Zn_ySe$, where a similar conclusion on sign-changing pairing was drawn[26].

The selection of the integrated range is based on the following considerations. First, signals at small $q$ are not included as they are usually due to disorder. Second, the integrated regions were carefully determined based on the topography of the Fermi surface. As it is difficult to distinguish $\mathbf{q_1}$ from different scattering channels, the area is selected so that the total signal due to inter- and intra-band large **q**-scattering are all well covered. The integral region is selected as $0.24/a_{Se} < \mathbf{q} < 0.64/a_{Se}$, slight smaller than that of $(Li_{1-x}Fe_x)OHFe_{1-y}Zn_ySe$ ($0.18/a_{Se} < \mathbf{q} < 0.77/a_{Se}$[35] or $0.37/a_{Se} < \mathbf{q} < 0.74/a_{Se}$[26]), due to the smaller electron pockets at M point for single layer FeSe/STO[36]. Calculations were done for three different integrated areas: $0.24\pi/a < \mathbf{q} < 0.64\pi/a$, $0.32\pi/a < \mathbf{q} < 0.64\pi/a$, and $0.24\pi/a < \mathbf{q} < 0.72\pi/a$, where similar positive and negative peaks are observed at the in-gap bound state energy, only with slight differences in intensity (Supplementary Fig. S15), further demonstrating that the appearances of these peaks are not due to artifacts such as integration areas.

Similar analysis was done for another type A defect with bound states located at ±6.4 meV (Supplementary Fig. S16), as well as a type B defect with bound states at ±9.0 meV (Supplementary Fig. S17), where corresponding positive and negative peaks were also observed, further supporting the sign-changing scenario.

## Discussion

Single layer FeSe/STO only consists of electron pockets at $M$ point and meanwhile it is fully gapped without nodes. Therefore, the three possible pairing symmetries are plain $s$-wave, nodeless

*d*-wave, and bonding-antibonding s-wave[20], as summarized in Supplementary Fig. S18. A plain *s*-wave has been suggested[13] based on quasiparticle interference off *adsorbates on the surface* of FeSe, where impurity-induced bound states were observed only for magnetic impurities. However, recent work clearly shows that even nonmagnetic surface adsorbates such as K and Pb can induce bound states near the superconducting gap edge[21,22], indicating sign-reversing pairing scenarios. In this work, in-gap bound states from a nonmagnetic impurity *within the Fe plane* is observed, which exhibit a strong spatial electron-hole asymmetry, also indicating a sign-changing pairing parameter. This is further confirmed by our phase-referenced defect bound state QPI data. Thus, the possibility of plain *s*-wave can be ruled out, which leaves two scenarios for the pairing symmetry of single layer FeSe/STO: *s*+– and nodeless *d*-wave. However, in both cases, scattering between Fermi surfaces of different orbital character with different signs are possible (Supplementary Fig. S19), hence cannot be distinguished by this method. Our recent work shows that one-dimensional scattering where the momentum **q** is always perpendicular to edges can provide a means to further differentiate these two pairing scenarios[37].

In conclusion, we systematically studied the in-gap bound states induced by nonmagnetic impurities within the Fe-plane in single layer FeSe/STO by STM/S. These bound states exhibit a strong spatial electron-hole asymmetry whose energy position is correlated with local doping. We further carry out phase-referenced defect bound state QPI, which reveals corresponding positive and negative peaks at the bound state energies, indicating a sign change for the order parameter. Our findings provide critical insight into the pairing symmetry of this intriguing high temperature superconductor single layer FeSe/STO.

**Methods**

***MBE growth and STM/S characterization:*** Single layer FeSe films were prepared by MBE under the Se-rich conditions using high purity Fe (99.995%) and Se (99.9999%) from an e-beam source and a K-cell on the Nb-doped (0.05 wt.%) $SrTiO_3$(001) substrate at 400 °C. The as-grown FeSe films were capped with ~20 nm amorphous Se films, and then transferred in air to a Unisoku-1300 STM system (air exposure time <15 min), where they were annealed at 200 °C for 2 hrs first to remove the extra Se and subsequently at between 460 and 480 °C for 1 h to reach superconductivity. Low-temperature STM/S measurements were carried out using a polycrystalline PtIr tip. The STS was acquired by using a lock-in technique with a bias modulation $V_{rms}$=0.5 mV at 732 Hz, unless otherwise specified.

***Defect bound state QPI***[24–27]***:*** For E>0, the phase referenced QPI signals $g_{pr}(\mathbf{q}, E)$ are expressed as

$$g_{pr}(\mathbf{q}, +E) = |g(\mathbf{q}, +E)| \times Re\left[e^{i(\theta_{\mathbf{q},+E} - \theta_{\mathbf{q},+E})}\right] = |g(\mathbf{q}, +E)| \quad (1)$$

$$g_{pr}(\mathbf{q}, -E) = |g(\mathbf{q}, -E)| \times Re\left[e^{i(\theta_{\mathbf{q},-E} - \theta_{\mathbf{q},+E})}\right] = |g(\mathbf{q}, -E)| \cos(\theta_{\mathbf{q},-E} - \theta_{\mathbf{q},+E}) \quad (2)$$

The phase term can be expressed as

$$\cos(\theta_{\mathbf{q},-E} - \theta_{\mathbf{q},+E}) = \cos\theta_{\mathbf{q},-E}\cos\theta_{\mathbf{q},+E} + \sin\theta_{\mathbf{q},-E}\sin\theta_{\mathbf{q},+E}$$

$$= \frac{Re[g(\mathbf{q},-E)]}{|g(\mathbf{q},-E)|}\frac{Re[g(\mathbf{q},+E)]}{|g(\mathbf{q},+E)|} + \frac{Im[g(\mathbf{q},-E)]}{|g(\mathbf{q},-E)|}\frac{Im[g(\mathbf{q},+E)]}{|g(\mathbf{q},+E)|} \quad (3)$$

Here, *Re* denotes the real part of $g(\mathbf{q}, E)$ while *Im* denotes the imaginary part. Based on Eq. (1), $g_{pr}(\mathbf{q}, +E)$ is always positive at positive energy, while at negative energy $g_{pr}(\mathbf{q}, -E)$ is determined by the phase term $\cos(\theta_{\mathbf{q},-E} - \theta_{\mathbf{q},+E})$. If there is a sign change between the initial and final state, the phase shift would be π, hence inducing a negative value since $\cos(\theta_{\mathbf{q},-E} - \theta_{\mathbf{q},+E}) = \cos(\pi) = -1$.

**Data availability:** The data supporting the findings of this study are available within the paper, its Supplementary Information files, and from the corresponding author upon reasonable request.

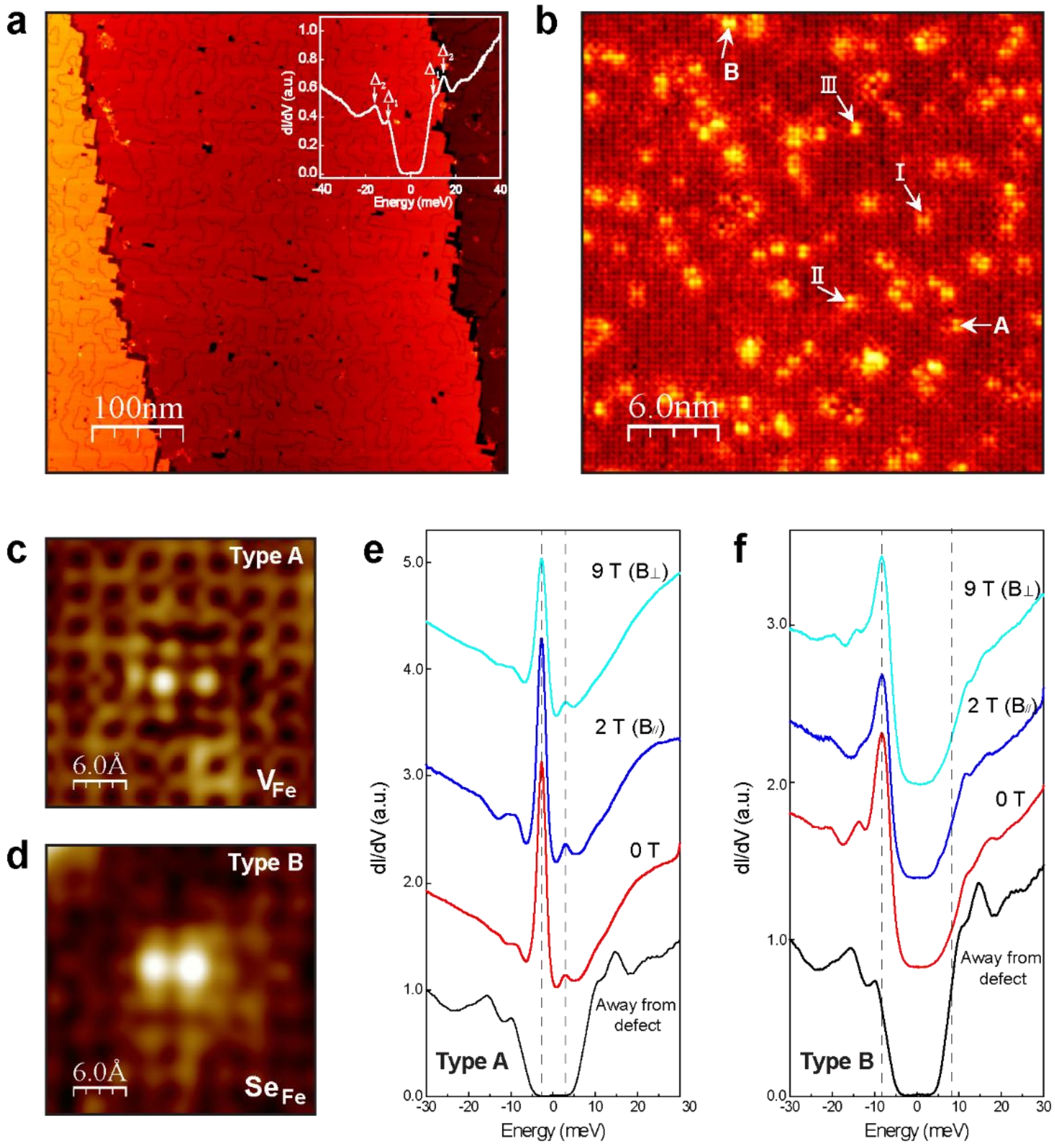

**Figure 1 Surface morphology and defects for single layer FeSe/STO films. a**, STM topographic image of single layer FeSe/STO films. Inset: A typical dI/dV spectrum measured at 4.3 K away from defects. The two-gap feature is labeled as $\Delta_1$ and $\Delta_2$. **b**, Atomic resolution image of various intrinsic defects. Setpoint: $V$=20 mV, $I$=500 pA. **c-d**, Close-up atomic resolution images of the type A and B defect labeled in **b**. Setpoint: $V$=30 meV, $I$=200 pA. **e-f**, Magnetic-field-dependent tunneling spectra for type A and B defects, respectively. The peak positions of the impurity bound states do not shift or split under an out-of-plane 9 T ($B_{\perp}$) or in-plane 2 T ($B_{//}$) magnetic fields, indicating their non-magnetic nature. The black spectra are measured at regions away from defects,

demonstrating a typical double-gap structure. The spectra labeled as 0 T, 2 T ($B_{//}$) and 9 T ($B_{\perp}$) are collected at the defect center in **c** and **d**, which are offset vertically for clarity.

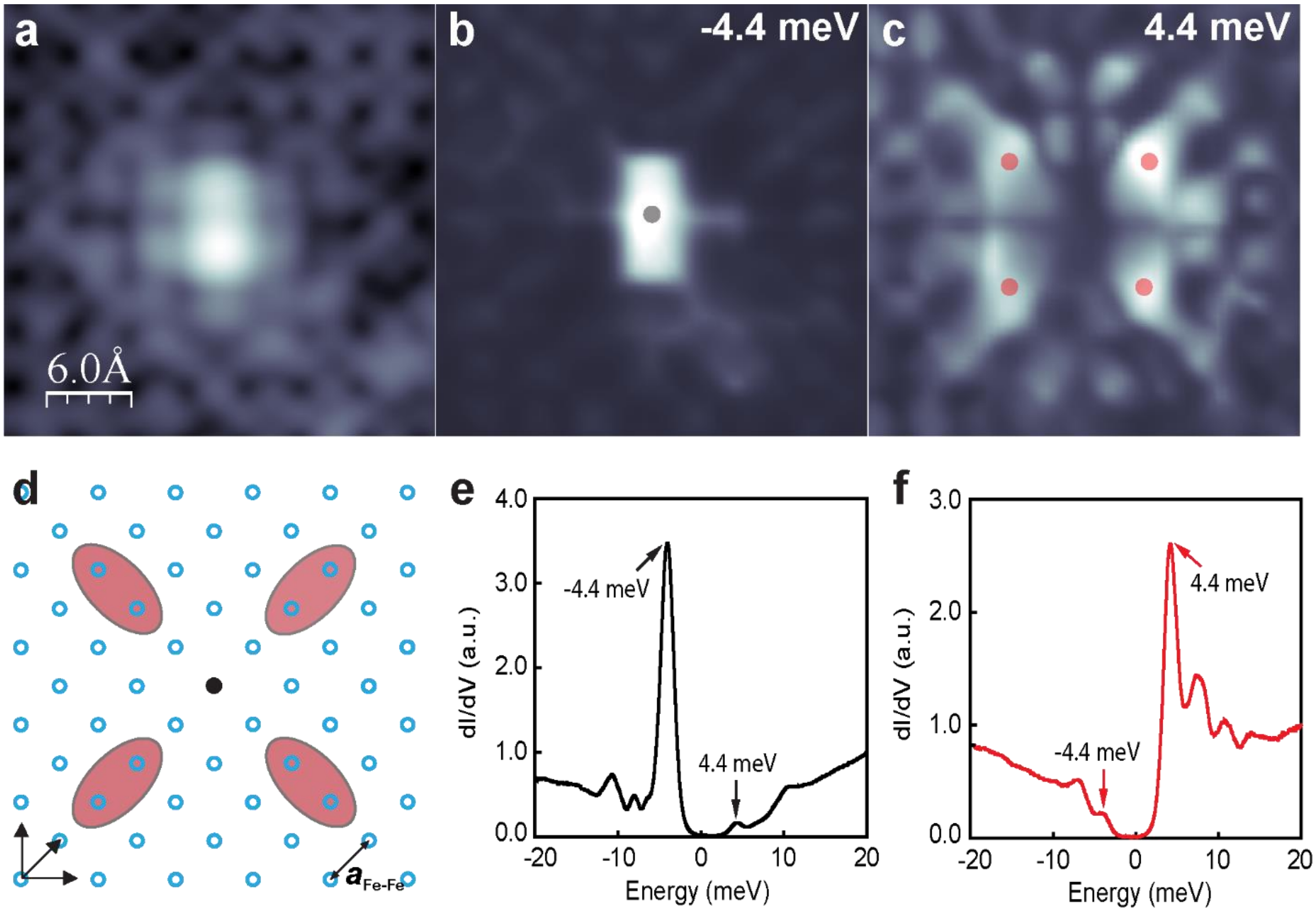

**Figure 2 Spatial distribution of bound states for dumbbell defect A (Fe vacancy). a**, Topographic image for Type A defect. Setpoint: $V$=–30 mV, $I$=1000 pA. **b-c**, Differential conductance maps at –4.4 meV and 4.4 meV, respectively, over the same region in **a**. Setpoint: $V$=12 meV, $I$=500 pA. **d**, Schematic representation of the Fe lattice. The distribution of the bound state at positive energy is illustrated by 4-fold red ellipses based on conductance map in **c**. The Fe vacancy defect is labeled by solid black dot in the center. **e**, dI/dV spectrum measured at the defect center (at the grey dot in **b**). **f**, dI/dV spectrum taken at one of the four bright lobes as marked by red dots in **c**.

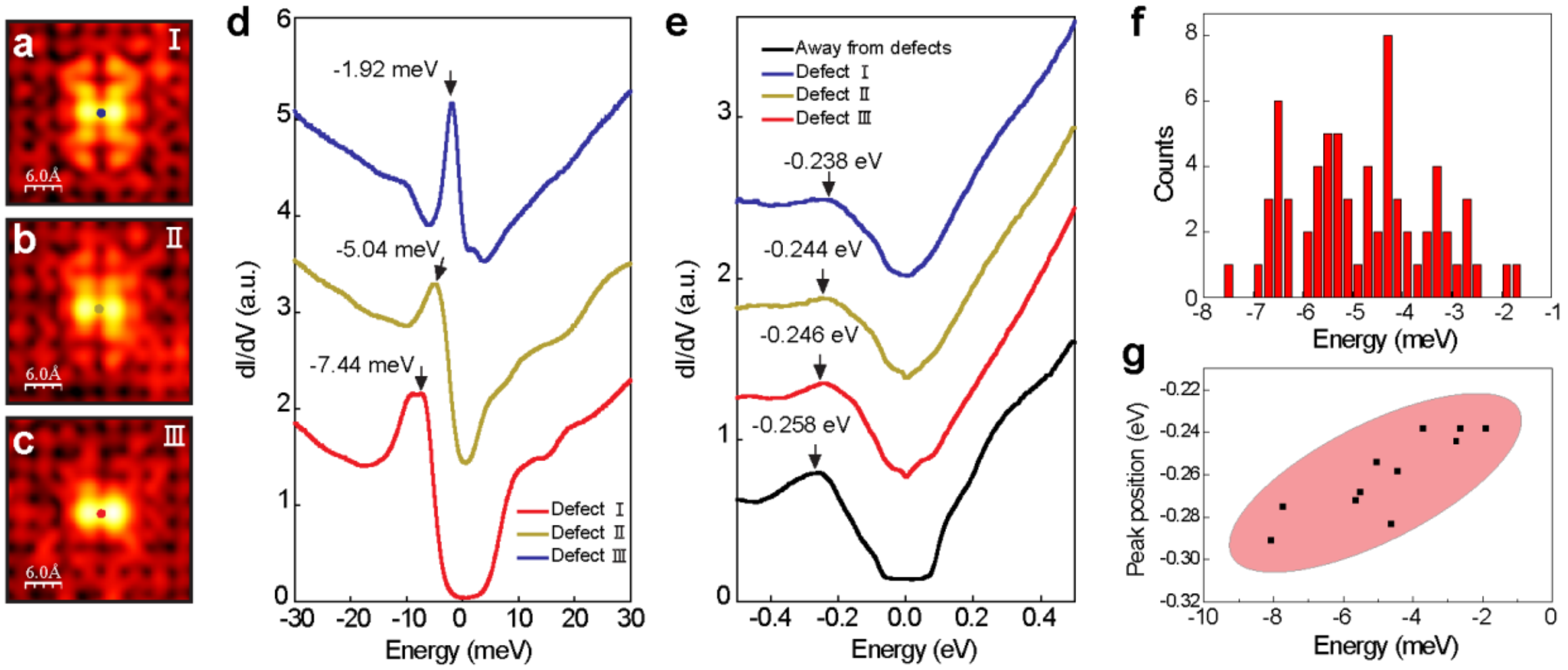

**Figure 3 Correlation between bound state energy and local doping level near dumbbell defects. a-c**, Topographic images of three Type A defects on single layer FeSe/STO films, labeled asI, II, III in Fig. 1**b**. Setpoint: *V*=20 mV, *I*=1.0 nA. **d**, Tunneling spectra showing bound states for defectsI(-1.92 meV), II(-5.04 meV), and III(-7.44 meV). **e**, Larger energy range tunneling spectra for defects I, II, III, and regions away from the defects, showing that the impurity bound state moves closer to the superconducting edge with more electron doping. **f**, Histogram of energies of bound states based on measurements from 70 dumbbell defects. **g**, Correlation between the normal-state and bound-state energies.

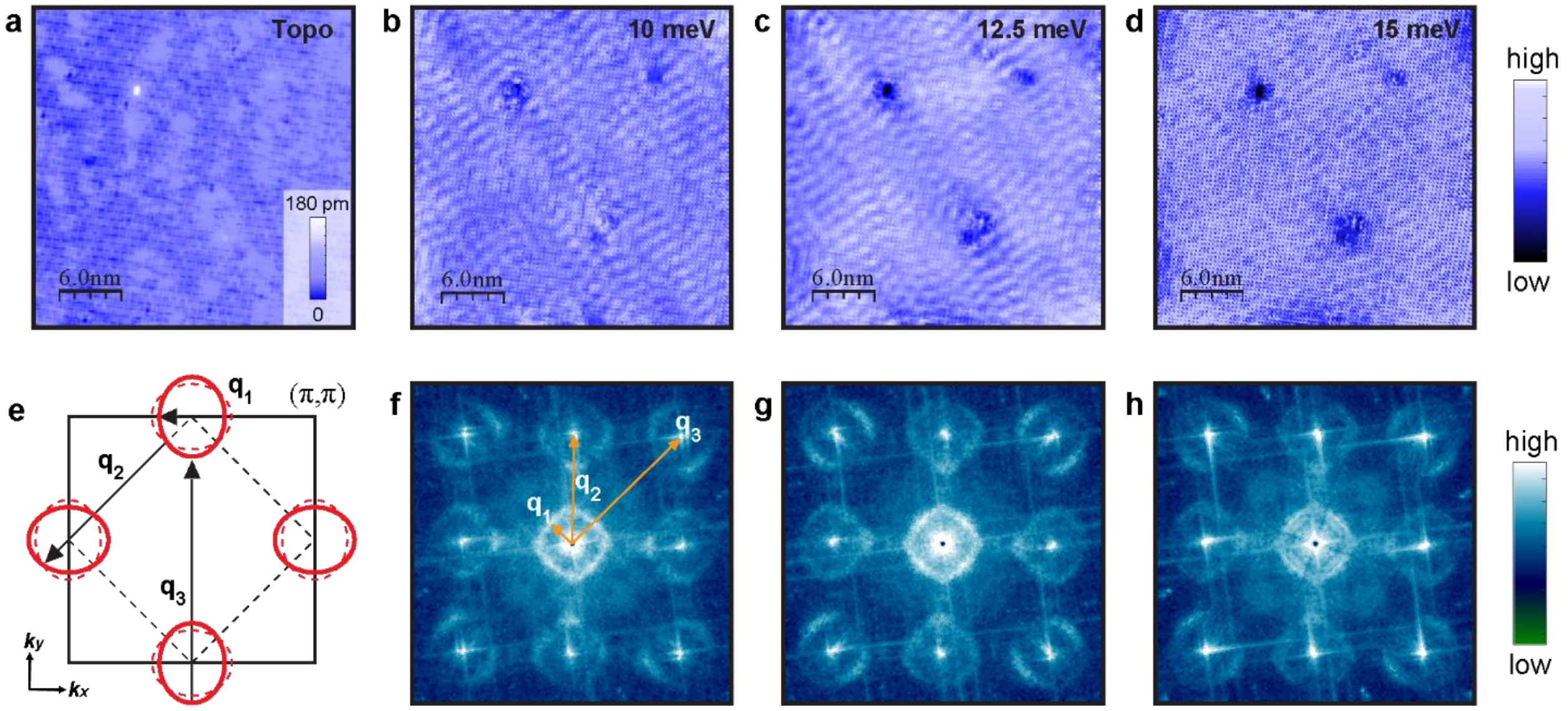

**Figure 4 Differential conductance maps in real space and corresponding quasiparticle interference patterns. a,** Topographic image of a 30 × 30 $nm^2$ region. Setpoint: $V$=16 mV, $I$=0.1 nA. **b-d,** Real space differential conductance maps $g(\mathbf{r}, E)$ in **a** (setpoint: $V$=16 mV, $I$=0.1 nA, $V_{rms}$=1 mV). **e,** Schematic of Fermi surface and Fermi pocket scatterings of single layer FeSe/STO in $k$ space. The solid square is the unfolded Brillouin zone (BZ) and the dashed square is the folded BZ. The red solid and dashed ellipses are the electron and folded pockets at BZ corners. The black arrows denote possible scattering vectors $\mathbf{q}_1$, $\mathbf{q}_2$, and $\mathbf{q}_3$. **f-h,** Fast Fourier transformation (FFT) of real-space differential conductance maps in **b-d**, respectively. All the FFTs are four-fold symmetrized and the core center is subtracted by a 2D-gaussian-function background.

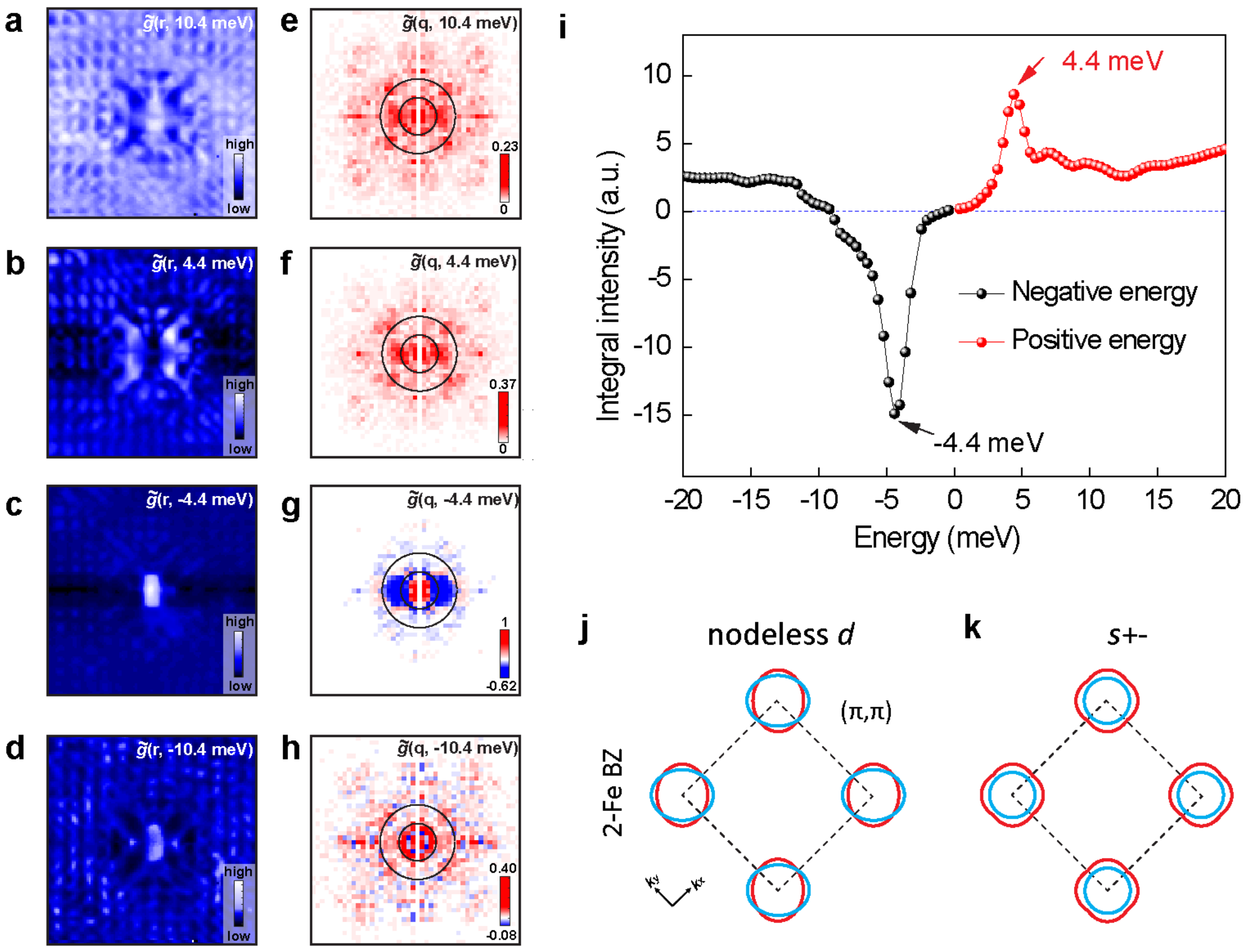

**Figure 5 Phase referenced defect bound state QPI for single Fe vacancy defect. a-d**, Real space differential conductance maps $g$ (**r**, $E$) centered at the Fe vacancy site at 10.4, 4.4, –4.4, and –10.4 meV with single Fe vacancy in the center. Set point: $V$=12 mV, $I$=0.5 nA, $V_{rms}$=0.5 mV. **e-h**, Corresponding phase-referenced DBS-QPI patterns $g_{pr}(\mathbf{q}, E)$ at the same energies. **i**, Integrated signals of $g_{pr}(\mathbf{q}, E)$ as a function of energy. The integrated region is chosen between the two black circles as shown in (**e-h**) with the range of $0.24\pi/a_{Se\text{-}Se} < \mathbf{q} < 0.64\pi/a_{Se\text{-}Se}$. The peaks at $-E_B$ and $+E_B$ marked by black and red arrows coincides with the impurity-induced bound states in Figs. 2**e-f**. **j-k**, Possible pairing symmetry in single layer FeSe/STO: nodeless $d$-wave pairing (**j**) and $s_{+-}$-wave pairing (**k**). The signs of superconducting gap are shown in red and blue.

**Acknowledgements**

Research supported by the U.S. Department of Energy, Office of Basic Energy Sciences, Division of Materials Sciences and Engineering under Award DE-SC0017632.

**Competing interests:** The authors declare no competing interests.

**Author contributions**

Huimin Zhang and Lian Li conceived the research. Zhuozhi Ge prepared single layer FeSe/STO films by MBE. Huimin Zhang carried out the low temperature STM/S measurements. Michael Weinert carried the calculations. Huimin Zhang and Lian Li analyzed the data and wrote the paper. All authors discussed the results and commented on the manuscript.